\documentclass[12pt]{article}
\usepackage{graphicx}
\begin{document} 

\title{The domino effect for markets}

\author{Christian Schulze\\
Institute for Theoretical Physics, Cologne University\\D-50923 K\"oln, Euroland}

\maketitle

e-mail: ab127@uni-koeln.de
\bigskip

Abstract:
A generalization of the Cont-Bouchaud market model to three markets agrees with 
the correlations netween New York, Tokyo, and Frankfurt observed by Vandewalle
et al.


\vspace*{1cm}
As studied by Vandewalle et al \cite{liege}, different stock markets like 
Tokyo, Frankfurt and New York are not independent of each other. We try to 
reproduce this observation in the well-known Cont-Bouchaud \cite{cb,rev}
percolation model. We will find that indeed in about 32 \% of the cases all
three markets have the same sign of change, as seen in reality \cite{liege}.

The Cont-Bouchaud model\cite{cb,rev} treats each percolation cluster as 
a company of individuals, who either all buy together (with probability
$a_{buy}$), or all sell together (with probability $1-a_{sell}$), or all
do not act (with probability $1-a_{buy}-a_{sell}$). For $a_{sell}=a_{buy}$
the market goes up or down with equal probability since supply and demand
agree on average. Thus if then three markets are simulated together, in 1/8
of the cases one has all three of them going up, in another 1/8 all of them go
down, and in the remaining 3/4 of cases the results are mixed. In reality,
however, in 17 \% of the trading days, Tokyo, Frankfurt and New York went up 
together, in another 17 \% they went down together, and in only 66 \% the 
results were mixed. Obviously, the results of the three markets influence each
other, facilitated by the rotation of the Earth.

{\begin{figure}[htb]
\includegraphics[angle=-90,scale=0.5]{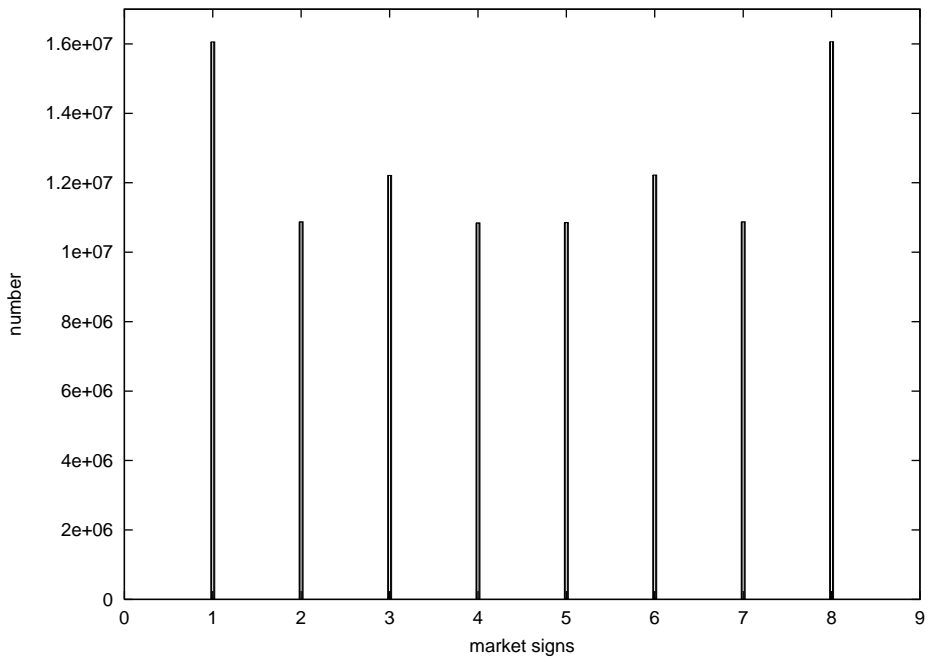}
\caption {Histogram from 100,000 simulations with 1000 iterations each
to have one of the eight possible 
combinations $S_1S_2S_3$ of signs $S_i = {\rm sign}(r_i)$, with  $---, --+, -+-,
\dots, +++$ from left to right. Very similar results were published for reality
\protect\cite{liege}.}
\end{figure}

We thus assume 
$$ a_{buy,i} = 0.05(1  + \sum_k c_{k,i} r_k); \quad 
a_{sell,i} = 0.05(1  - \sum_k c_{k,i} r_k)$$
with $k=i$ omitted: 1 = New York, 2 = Tokyo, 3 = Frankfurt.
Here $c_{i,k}$ gives the influence of market $i$ on market $k$, and the 
return $r_i$ is the relative price change of the market (Dow Jones 
Industrials, Nikkei, Dax) of the preceding day. We take $c_{i,k}=0$ except for
$c_{1,2}=0.001, \; 
c_{1,3}=0.002, \; c_{2,3}=0.0005$ for a $101 \times 101$ Ising lattice
at $T/T_c = 1.1$ as in \cite{silva}. The figure shows a probability of about 32
\% for all three $r_i$ to have the same sign $S_i={\rm sign}(r_i)$. For uncorrelated
markets this probability would only be 1/4, while in reality it is 34 \%. 
Taking appreciably different coupling coefficients $c_{ik}$ gives different 
results, deviating from reality.

In conclusion, we found excellent agreement but do not exclude that the same
agreement would also be found from other market models.

We thank Deutsche Forschungsgemeinschaft for support and D. Stauffer for help 
with the manuscript.

\end{document}